\documentclass[twocolumn]{aastex63}

\usepackage{txfonts}
\usepackage{tasks}
\usepackage{tabularx}
\usepackage{float}
\usepackage{graphicx}
\usepackage{rotating}
\usepackage{booktabs}
\usepackage{dblfloatfix}
\usepackage{afterpage}
\graphicspath{ {images/} }
\newcolumntype{Y}{>{\centering\arraybackslash}X}

\received{2019 August 9}
\revised{2019 October 22}
\accepted{2019 October 23}

\submitjournal{ApJ Letters}

\begin{document}

\title{Evidence of a substellar companion to AB Dor C}

\correspondingauthor{J.B. Climent}
\email{j.bautista.climent@uv.es}

\author{J.B. Climent}
\affiliation{Departament d'Astronomia i Astrof\'isica, Universitat de Val\`encia, C. Dr. Moliner 50, 46100 Burjassot, Val\`encia, Spain}

\author{J.P. Berger}
\affiliation{Univ. Grenoble Alpes, CNRS, IPAG, 38000 Grenoble, France}

\author{J. C. Guirado}
\affiliation{Departament d'Astronomia i Astrof\'isica, Universitat de Val\`encia, C. Dr. Moliner 50, 46100 Burjassot, Val\`encia, Spain}
\affiliation{Observatori Astron\`omic, Universitat de Val\`encia, Parc Cient\'ific, C. Catedr\'atico Jos\'e Beltr\'an 2, 46980 Paterna, Val\`encia, Spain}

\author{J.M. Marcaide}\altaffiliation{Visiting Professor at the Department of Quantum Physics and Astrophysics, and the Institute of Cosmos Sciences of the University of Barcelona}
\affiliation{Departament d'Astronomia i Astrof\'isica, Universitat de Val\`encia, C. Dr. Moliner 50, 46100 Burjassot, Val\`encia, Spain}

\author{I. Mart\'{\i}-Vidal}
\affiliation{Departament d'Astronomia i Astrof\'isica, Universitat de Val\`encia, C. Dr. Moliner 50, 46100 Burjassot, Val\`encia, Spain}
\affiliation{Observatori Astron\`omic, Universitat de Val\`encia, Parc Cient\'ific, C. Catedr\'atico Jos\'e Beltr\'an 2, 46980 Paterna, Val\`encia, Spain}

\author{A. M\'erand}
\affiliation{European Southern Observatory, Karl-Schwarzschild-Str. 2, 85748 Garching bei M\"unchen, Germany}

\author{E. Tognelli}
\affiliation{Department of Physics E. Fermi, University of Pisa, Largo Bruno Pontecorvo 3, 56127, Pisa, Italy}

\author{M. Wittkowski}
\affiliation{European Southern Observatory, Karl-Schwarzschild-Str. 2, 85748 Garching bei M\"unchen, Germany}

\begin{abstract}

Studies of fundamental parameters of very low-mass objects are indispensable to provide tests of stellar evolution models that are used to derive theoretical masses of brown dwarfs and planets. However, only objects with dynamically determined masses and precise photometry can effectively evaluate the predictions of stellar models. AB Dor C (0.090 M$_{\rm\odot}$) has become a prime benchmark for calibration of theoretical evolutionary models of low-mass young stars. 
One of the ambiguities remaining in AB Dor C is the possible binary nature of this star.
We observed AB Dor C with the VLTI/AMBER instrument in low-resolution mode at the J, H and K bands. The interferometric observables at the K-band are compatible with a binary brown dwarf system with tentative components AB Dor Ca/Cb with a K-band flux ratio of 5\%~$\pm$~1\% and a separation of 38~$\pm$~1 mas. This implies theoretical masses of 0.072~$\pm$~0.013 M$_{\rm \odot}$ and 0.013~$\pm$~0.001 M$_{\rm \odot}$ for each component, 
near the hydrogen-burning limit for AB Dor Ca, and near the deuterium-burning limit, straddling the boundary between brown dwarfs and giant planets, for AB Dor Cb.
The possible binarity of AB Dor C alleviates the disagreement between observed magnitudes and theoretical mass-luminosity relationships. 

\end{abstract}


\section{Introduction} \label{sec:intro}

Stellar evolution models are essential to infer star fundamental parameters such as radius, mass, and/or age. 
Their reliability has long been tested and validated by the general good agreement between predictions and measurements. However, only recently have accurate measurements of stellar masses and radii become accessible in the case of low and very low-mass stars, thus allowing more stringent tests on stellar models. In the particular case of pre-main-sequence (PMS) stars, the models show an increasing difficulty in accurately reproducing some of the characteristics of star with masses below 1.2 \(M_\odot\) \citep[see, e.g.,][]{2004ApJ...604..741H,2012MNRAS.420..986G,2014NewAR..60....1S}.  

Only stellar systems with  dynamically determined masses can effectively be used to test and check the predictions of the  models 
(see recent works of \citet{2017ApJS..231...15D} and \citet{2019ApJ...871...63M}). AB Doradus (AB Dor) represents one such case. It is a PMS quadruple system formed by two pairs of stars separated by 9$''$, AB Dor A/C and AB Dor Ba/Bb \citep{2005Natur.433..286C,2006A&A...446..733G}, giving name to the AB Doradus moving group (AB Dor-MG). The main star of this system, the K0 dwarf AB Dor A ($K_s$ = 4.686) has been extensively studied at all wavelengths, from the UV to radio \citep{2002MNRAS.332..409G,1997ApJ...490..835G}. Precise Hipparcos and 
very long baseline interferometry 
(VLBI) observations provided an accurate distance measurement ($d=15.06\pm0.07$ pc) and revealed the presence of AB Dor C, a low-mass companion with 0.090 \(M_\odot\), orbiting AB Dor A at an average angular distance of 0.2$''$ \citep{1997ApJ...490..835G}. The pair AB Dor A/C has also been observed by different near-infrared instruments at the VLT \citep{2005Natur.433..286C,2007ApJ...665..736C,2008A&A...482..939B} allowing independent photometry of AB Dor C ($K_s$ = 9.5) which, along with the dynamical mass determination, served as a benchmark for young, low-mass stellar evolutionary models \citep[and references therein]{2017A&A...607A..10A}. 

Previous comparisons of observed magnitudes with theoretical mass-luminosity relationships suggested that the models tend to underpredict the mass of AB Dor C or, equivalently, overpredict the flux of the object, especially at the $J$ and $H$ bands \citep{2005Natur.433..286C}. 
This disagreement was also noted in studies of the other pair of the system, AB Dor Ba/Bb \citep{2014A&A...570A..95W,2018A&A...620A..33J}. The authors argued that theoretical models tend to be consistent in the case of young moving groups but not in older associations such as the AB Dor moving group.
This tendency was reinforced by the study of other members of this moving group, such as  GJ 2060 AB \citep{2018A&A...618A..23R} or LSPM J1314+1320 AB \citep{2016ApJ...827...23D}. In the case of AB Dor C,
most of the difficulty in validating the model predictions comes from the uncertainty in age and the possible binary nature of the object. Regarding the latter, \citet{2005astro.ph..2382M} pointed out that if AB Dor C were a binary brown dwarf, the overluminosity shown by the models, including the permanent disagreement in the $J$ and $H$ filters, could easily be corrected assuming reasonable mass ratios. Indeed, the determination of the possible binary nature of AB Dor C is an important issue for an object acting as calibrator of young, low mass objects that needs to be addressed.

In this Letter we present interferometric evidence of the presence of a low-mass companion to AB Dor C from VLT Interferometer (VLTI) observations performed with the Astronomical Multi-BEam combineR (AMBER) focal instrument \citep{2007A&A...464....1P}, installed at the ESO facilities in Cerro Paranal, Chile. Attempts to observe this object with the GRAVITY instrument are also reported.  We describe the observations and data reduction in Section 2. The analysis is presented in Section 3 and the comparison with stellar models and discussion are presented in Section 4. Finally, in Section 5 we present our conclusions.\\
  
\section{Observations}

   \begin{figure}[h]
    \centering
    \includegraphics[width=\linewidth]{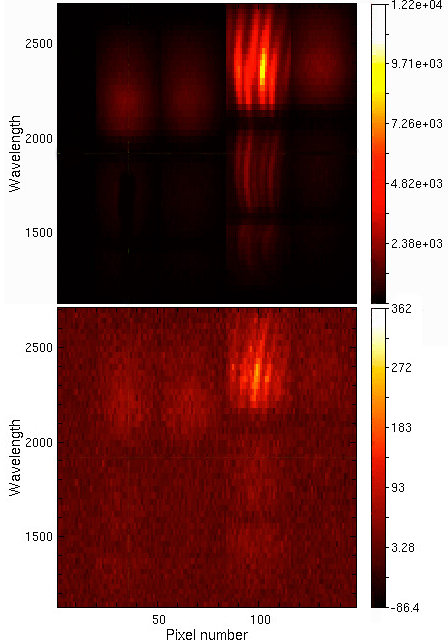}
    \caption{
    AMBER raw detector image (single frame) obtained with the triplet UT1-UT2-UT4 in low-resolution mode on AB Dor A (upper plot) and AB Dor C (lower plot), the latter taken with a non-standard, off-axis fringe-tracking configuration (see text). The upper (lower) half of the plots corresponds to $K$ ($H$) band. From left to right, the first, second, and fourth columns represent the photometric beams for each one of the three telescopes, while the third column contains the interferometric signal. Notice the clear K-band detection on AB Dor C. The ratio between the intensity of both interferometric channels roughly indicates the flux ratio between AB Dor A and C.
    }
    \label{fig:rawimage}
    \end{figure}

    \begin{table}
    \centering
    \caption{Observation log of AB Dor C and calibrators.}
    \label{table:logs}
    \begin{tabular}{ccccc} 
    \hline\hline
    Obs. Time & Target & Triplet & Mode & Seeing\\
    \hline
    28/12/2012 02:40 & HD 35199 & U1-U2-U4 & Low JHK & 0.79''\\
    28/12/2012 03:17 & AB Dor C & U1-U2-U4 & Low JHK & 0.63''\\
    28/12/2012 03:43 & AB Dor C & U1-U2-U4 & Low JHK & 0.72''\\
    28/12/2012 04:17 & AB Dor C & U1-U2-U4 & Low JHK & 0.74''\\
    28/12/2012 04:40 & AB Dor A & U1-U2-U4 & Low JHK & 0.74''\\
    \hline
    \end{tabular}
    \begin{flushleft}
    \footnotesize{\textbf{Note.} Due to the nonstandard observing configuration (see the text), HD 35199 could only be observed at the beginning of the observation.}
    \end{flushleft}
    \end{table}
    
    \begin{figure}[h]
    \centering
    \includegraphics[width=\linewidth]{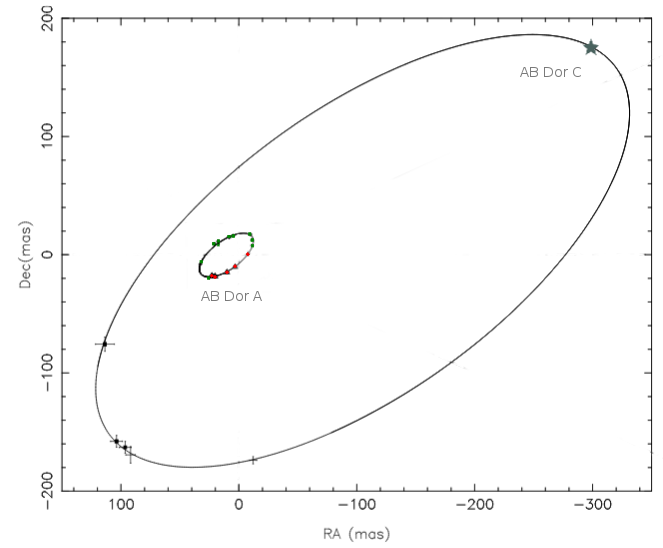}
    \caption{Absolute orbits of AB Dor A and AB Dor C, adapted from \citep{2017A&A...607A..10A}. The map is centered at the center of mass of the system. Measured positions of AB Dor A are marked with red (Hipparcos data) and green (VLBI data) dots. Previous VLT/NACO measurements of AB Dor C are plotted with points while the expected position of AB Dor C at the time of our observation (2012.9918) is marked by a star.
    }
    \label{fig:orbit}
    \end{figure}

The observations of AB Dor C were performed with the VLTI using the AMBER instrument with the external fringe tracker FINITO (Fringe-Tracking Instrument of NIce and TOrino) in low resolution mode at the J, H and K bands (programme 090.C-0559(A)). However, due to a technical problem, the J-band did not perform well and was not used in our analysis. The H-band was tested but finally also discarded in the analysis, as discussed below.
The observations were performed on 2012 December 28 from 02:40 to 04:40 UT using the 8.2 m unit telescopes (UT) with the configuration UT1-UT2-UT4. 
Due to the faint magnitude of AB Dor C ($K_s$ = 9.5), we used a nonstandard observing configuration that consisted of using AB Dor A as a fringe tracker to increase the integration time on AB Dor C. 
To achieve this, first we set
AMBER in low-resolution mode with a DIT of
0.1 s; second, we found and locked the fringes
of AB Dor A in the fringe tracker FINITO; and
third, we offset AMBER (through tip/tilt correction) to find the fringes of AB Dor C. This
“off-axis” fringe tracking allowed an exposure
time on AB Dor C longer than that imposed by
the atmospheric piston. Fringes were seen in every single frame (see Fig. \ref{fig:rawimage}), which could be properly averaged to obtain the visibility data.
The procedure above benefited
from (i) a precise knowledge of the orbit of
AB Dor C \citep{2006A&A...446..733G}, which allowed
us to predict with milliarcsecond precision the
relative position AB Dor A/C; (ii) an optimum observing epoch, 2012 December, with
AB Dor C near apoastron (0.42" separated from A, see Fig. \ref{fig:orbit}), thus minimizing the
possible contamination from the brighter star
AB Dor A; and (iii) good
atmospheric conditions with a seeing of $\sim$0.7".
The star HD 35199 \citep[disk-equivalent diameter of 0.86 mas;][]{2006A&A...447..783M} was also observed to calibrate
the AB Dor C visibilities. The logs of these observations are shown in Table \ref{table:logs}.
    
In addition to the AMBER data, 4 hr of VLTI/GRAVITY time were allocated (program 0102.C-0297, with the telescopes UT1-UT2-UT3-UT4) and scheduled on 2017 December 9 to confirm our findings. However, the proximity of the much brighter AB Dor A (located at 0.2" from C) during the observing epoch prevented AB Dor C to be properly identified in the GRAVITY acquisition camera, therefore making the observation technically unfeasible. 

\section{Data reduction and analysis}\label{sect:analysis}
    
We obtained the raw visibility data using the software package amdlib v.3.0.8\footnote{The AMBER reduction package amdlib is available at: \url{http://www.jmmc.fr/data_processing_amber.htm}} \citep{2007A&A...464...29T,2009A&A...502..705C}. We selected and averaged the resulting visibilities of each frame using different criteria for the baseline flux and for the fringe signal-to-noise ratio (S/N; for more information see the AMBER Data Reduction Software User Manual\footnote{\url{http://www.jmmc.fr/doc/approved/JMMC-MAN-2720-0001.pdf}}. In particular, (i) we selected frames having a baseline flux with S/N larger than  1, 2, 3, 4, 5, 6, 7, 10, 15, and 20; (ii) for each of these selections, we kept the 5\%, 10\%, 20\%, 30\%, 40\%, 50\%, 60\%, 70\%, 80\%, and 100\% of the remaining frames with the highest fringe S/N, which effectively created a grid of 10 x 10 reduced data sets with different selection criteria; (iii) we made extensive tests to determine the robustness and consistency of each one of the data sets above (basically, we compared each data set to simulated visibilities of different source geometries, discarding those data sets producing unacceptable fits); and (iv) based on the previous tests, we selected the data set containing 5\% of the frames with highest fringe S/N chosen from those with a baseline flux S/N larger than 6. The calibration of the transfer function was made using the calibrator star HD 35199.

    \begin{figure}
    \centering
    \includegraphics[width=1\linewidth]{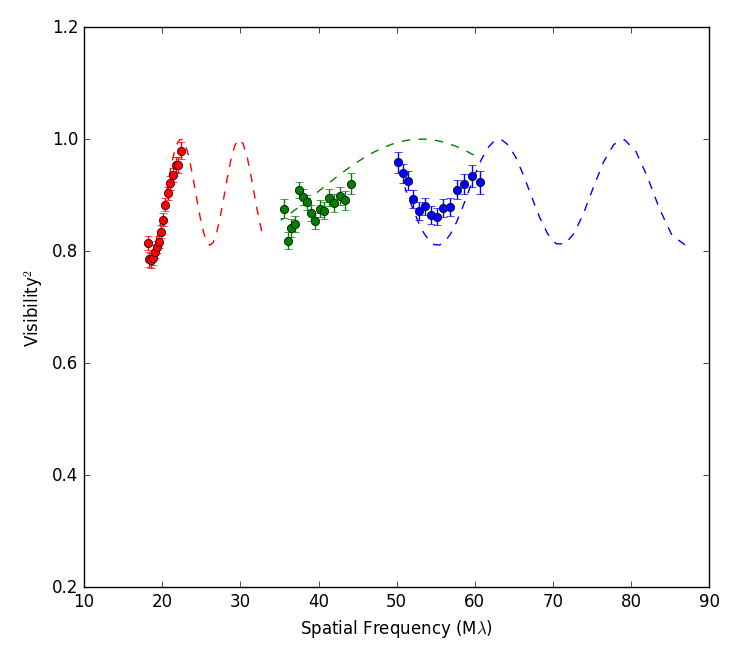}
    \includegraphics[width=1\linewidth]{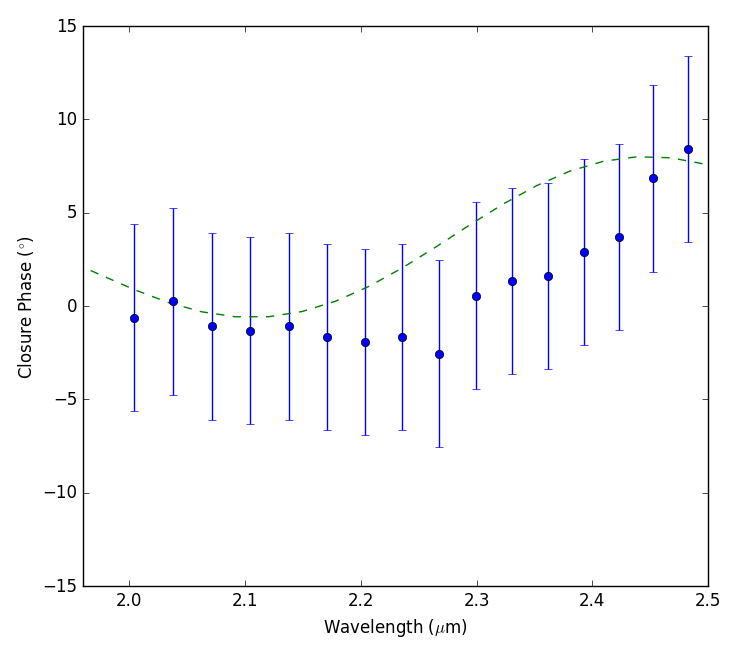}
    \caption{ (\textit{upper}): AMBER visibilities of AB Dor C. Different colors represent different baselines. The K-band observational data (circles) are best fitted by a binary system with the properties given in Table \ref{table:amberparameters} with the CANDID K-band method and whose visibilities are plotted in colored discontinuous lines. (\textit{lower}): AMBER K-band closure phases of AB Dor C. Blue dots correspond to observational data, while the dashed line represents the theoretical closure phases of the model given above. Indeed, the deviation from 0$^{\circ}$ suggests that AB Dor C is not a point source and possesses a more complex
    structure, modeled here by a binary.}
    \label{fig:ambervis}
    \label{fig:amberphases}
    \end{figure}
    
The values of the squared visibilities panel (Fig \ref{fig:ambervis} left) are far from corresponding to those of a pointlike source; rather, they indicate either the presence of an extended structure around AB Dor C, and/or the binary nature of the object. Actually, the sinusoidal behavior seen in the visibilities is a typical signature of a binary system \citep[e.g.,][]{2009A&A...507..317M}. Supporting the previous statement, the closure phase (Fig.\ref{fig:amberphases} right) displays a nonnegligible departure from the null value, indicating that the contrast between binary components should be relatively high \citep{2003RPPh...66..789M}. The fact that the visibilities do not decrease with baseline suggests, in principle, that the components are not resolved.

    We performed an exhaustive and systematic search for companions to AB Dor C using the software CANDID\footnote{[C]ompanion [A]nalysis and [N]on-[D]etection in [I]nterferometric [D]ata, available at: \url{https://github.com/amerand/CANDID}} \citep{2015A&A...579A..68G}. CANDID performs a least-squares fit of both the companion position and flux ratio at each starting position of a 2D grid using the interferometric observables, the squared visibilities, and the closure phases. We first used CANDID with only K-band data, revealing a companion to AB Dor C at a level of 24$\sigma$ with separation and flux ratio detailed in Table \ref{table:amberparameters}, where the number of sigmas indicates how significant the binary model is compared to a single star and is computed using formula (8) in \citet{2015A&A...579A..68G}. In agreement with this, the calculation of the corresponding $\chi^{2}$ for both the single and binary scenarios yields a clear preference for the presence of a companion:  $\chi^{2}_{single}$ = 14.07 and $\chi^{2}_{binary}$ = 2.60.

On the other hand, the quality of the fit is degraded (from 24$\sigma$ to 16$\sigma$) when using both the K and H bands, adding, at least, another spurious solution. Moreover, no detection is found with CANDID when using only the H band. 
Given the results above, we  conservatively restricted our interferometric data set to K-band only.

    \begin{table}
    \centering
    \caption{Best-fitting binary-model parameters for AB Dor C}
    \label{table:amberparameters}
    \begin{tabular}{cccc} 
    \hline\hline
    Method & Flux ratio & Separation (mas) & P.A ($^{\circ}$)\\
    \hline
    CANDID K band & 0.054~$\pm$~0.004 & 38.1~$\pm$~0.2 & 178~$\pm$~1 \\
    LITpro & 0.05~$\pm$~0.01 & 39~$\pm$~1 & 177~$\pm$~1 \\
    \hline
    \end{tabular}
    \end{table}


To assess the validity of the CANDID results, we also fitted the observed visibilities with LITpro (Lyon Interferometric Tool prototype), developed by the Jean-Marie Mariotti Center \citep[JMMC][]{2008SPIE.7013E..1JT}. 
We used a simple two-point model to simulate the suspected binary nature of AB Dor C. 
In contrast to the CANDID procedure, LITpro does not perform a systematic search of the parameters; therefore, aiming at identifying the best model, we initialized the fitting program with different sets of values for the free parameters, namely: flux ratio, binary separation, and position angle. In practice, we explore the following parameter space around the CANDID position: flux ratio between 2\% and 8\% in steps of 0.5\%, separation between 25 and 50 mas in steps of 0.5 mas, and P.A. between 140$^{\circ}$ and 220$^{\circ}$ in steps of 0.5$^{\circ}$. 
We selected as plausible fits the range of $\chi^{2}$ values that correspond to a 95\% confidence interval. The results of this parameter search are
given in Table \ref{table:amberparameters}, and coincide, within uncertainties, with the position found with the CANDID software and the K-band, strengthening the validity of our binary hypothesis for AB Dor C.
Given the plausibility of the results obtained with this binary model (and, admittedly, to avoid a possible overinterpretation of the data), we did not explore more complicated geometries with LITpro (i.e. binary with disks or envelopes). \\

From the analysis above, we conclude that AB Dor C is a binary system with components separated by 38~$\pm$~1 mas, and a flux ratio at the K-band of 5\%~$\pm$~1\%, where the uncertainties have been conservatively enlarged to cover the results of both software.



\section{Results and discussion}\label{sect:results}

    \begin{figure}[h]
    \centering
    \includegraphics[width=\linewidth]{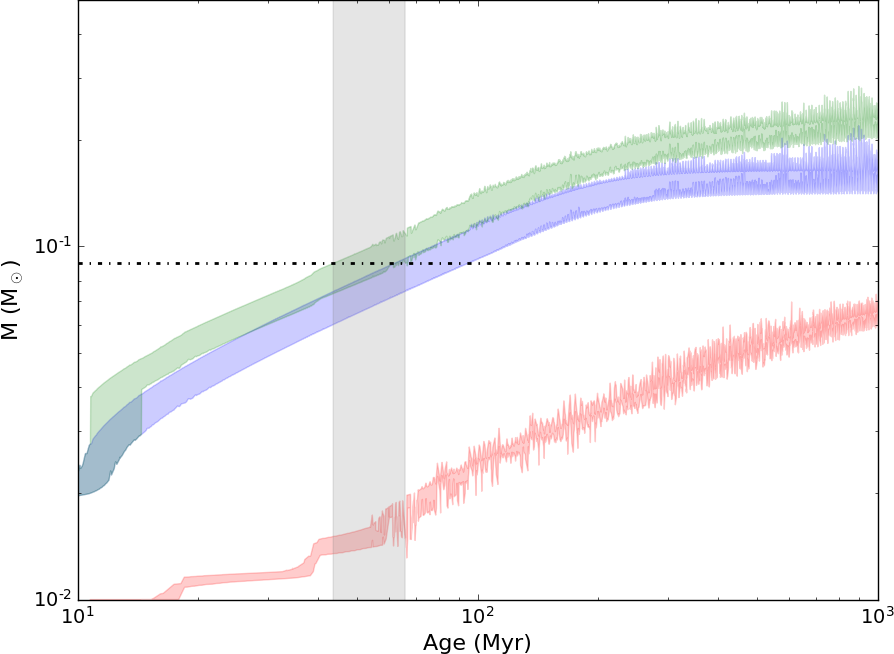}
    \caption{Mass versus age at constant $K_s$ magnitude: 8.43~$\pm$~0.16 (blue; Ca) and 11.7~$\pm$~0.3 (red; Cb) and the corresponding binary system (green; Ca+Cb). A dashed-dotted line represents the measured dynamical mass of 0.090 M$_{\rm \odot}$ \citep{2017A&A...607A..10A}, while the shadowed gray area represents the age range (44-66 Myr) resulting from the intersection of this measured dynamical mass with the measured Ca+Cb magnitude, according to the models by \citet{2018MNRAS.476...27T}.
    The irregularities in the model are more likely a product of the interpolation than a real physical effect. The use of the DUSTY+BHAC15 models produces very similar results.}
    \label{fig:mvstime}
    \end{figure}

The interferometric results presented allow us to characterize the components of the tentative binary in AB Dor C (AB Dor Ca/Cb). The combination of the flux ratio (Cb/Ca), 5\%~$\pm$~1\%, with the total (Ca+Cb) $K_s$ absolute magnitude of 8.38 $\pm$ 0.16 \citep{2008A&A...482..939B} implies a binary with magnitudes $K_s$ = 8.43~$\pm$~0.16 and $K_s$ = 11.7~$\pm$~0.3 for Ca and Cb, respectively.

\subsection{Comparison with evolutionary models}\label{sect:evolutionary_models}
 
With the individual magnitudes of the components, an estimate of the individual masses can be obtained by using PMS evolutionary models.
We computed the models using the Pisa stellar evolutionary code \citep{2018MNRAS.476...27T} for masses in the range 0.01--0.4 M$_{\rm \odot}$ and solar metallicity ([Fe/H] = 0). 
We also used  an interpolation of the DUSTY models 
\citep[][]{2000ApJ...542..464C}
and BHAC15 models 
\citep[][]{2015A&A...577A..42B}
to infer the dependence of the derived parameters on the adopted evolutionary tracks. The boundary between DUSTY and BHAC15 models was based on mass, with each model covering the ranges of 0.001--0.1 M$_{\rm \odot}$ for the former and 0.01--1.4 M$_{\rm \odot}$ for the latter.

Following the procedure described in Sect. 4.2 of \citet{2008ApJ...689..436L}, we used these evolutionary models and our inferred magnitudes of Ca and Cb to 
estimate the mass of each one of the components; the sum of these masses is the the total mass of the system, which is represented against the age in Fig. \ref{fig:mvstime}. 
The shadowed gray area indicates the possible combinations of masses for Ca and Cb accomplishing that the total mass of the system lies within the measured dynamical mass of 0.090$\pm$0.008 M$_{\rm \odot}$ for AB Dor C \citep{2017A&A...607A..10A}. This implies masses of 0.072~$\pm$~0.013 M$_{\rm \odot}$ and 0.013~$\pm$~0.001 M$_{\rm \odot}$ for each one of the components of the binary, which interestingly lie near the hydrogen-burning limit for the case of AB Dor Ca, and near the deuterium-burning limit, straddling the boundary between brown dwarfs and giant planets, for the case of AB Dor Cb.  
Given the relatively large mass ratio between AB Dor Ca/Cb, we notice 
that the tentative binarity of AB Dor C would not result in a substantial change in the age range when compared with previous estimates based on the same evolutionary models \citep{2017A&A...607A..10A}. Likewise, the presence of AB Dor Cb does not 
affect the age determinations based on model isochrones fitting to the members of the AB Dor moving group \citep{2015MNRAS.454..593B}.
  
\subsection{Binary hypothesis and photometry}

\begin{figure}
    \centering
    \includegraphics[width=\linewidth]{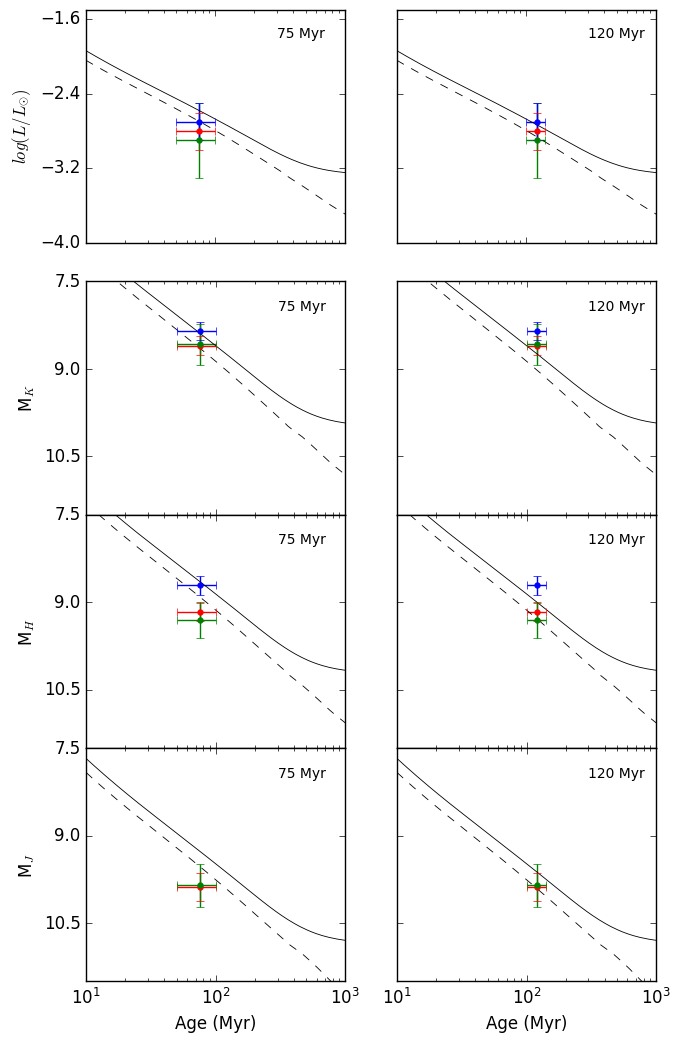}
    \caption{Evolutionary tracks of theoretical models \citep{2018MNRAS.476...27T} for a single object of 0.090 M$_{\rm \odot}$ (solid) and a binary system of 0.072~$\pm$~0.013 M$_{\rm \odot}$ and 0.013~$\pm$~0.001 M$_{\rm \odot}$ (dashed) compared with photometric measurements of \citet{2006ApJ...638..887L} (green dots), those of \citet{2007ApJ...665..736C} (red dots) and those of \citet{2008A&A...482..939B} (blue dots). The first row represents the bolometric luminosity derived from the photometric measurements and the bolometric corrections found in \citet{2015ApJ...810..158F}. The second, third, and fourth rows are for K, H, and J photometry, respectively. The left plots represent the first age scenario (75~$\pm$~25 Myr) and the right plots represent the second one (120~$\pm$~20 Myr). The DUSTY+BHAC15 models produce extremely similar results.}
    \label{fig:3bands}
    \end{figure}

How is the interpretation of published AB Dor C photometry affected by our binary hypothesis? In Fig. \ref{fig:3bands} we represent the cooling curves of AB Dor Ca/Cb \citep[models by][]{2018MNRAS.476...27T} compared with published AB Dor C photometric measurements \citep{2005Natur.433..286C,2006ApJ...638..887L,2007ApJ...665..736C,2008A&A...482..939B} for the two scenarios considered: 1) the (old) single-object scenario (with AB Dor C as a single object of 0.090 M$_{\rm \odot}$) and 2) the (new) binary scenario resulting from our detection (with AB Dor C as a binary with estimated masses of 0.072 M$_{\rm \odot}$ and 0.013 M$_{\rm \odot}$, according to the parameters given in Table 2). Published magnitudes at $J$, $H$, and $K$ bands are shown for the two most-considered age ranges in the literature: 75~$\pm$~25 Myr \citep{2007A&A...462..615J,2008A&A...482..939B} and 120~$\pm$~20 Myr \citep{2005ApJ...628L..69L,2007MNRAS.377..441O}, being the latter being favored by the recent works of \citet{2015MNRAS.454..593B} and \citet{2018ApJ...861L..13G}.

For the discussion that follows we use solely the fact that AB Dor C may be a binary system with masses 0.072 M$_{\rm \odot}$ and 0.013 M$_{\rm \odot}$ (as obtained in Sect. \ref{sect:evolutionary_models}) and previously reported photometric measurements.
As we can see in Fig. \ref{fig:3bands}, for an age of 75\,Myr (left column plots), the binary hypothesis produced an overall better agreement considering all the three bands than the single-object hypothesis. In fact, the binary tracks are compatible (to within 1.2$\sigma$ of the showed uncertainties), with all the photometric measurements, slightly favoring those reported by \citet{2007ApJ...665..736C} and \citet{2006ApJ...638..887L}. Turning to the age of 120\,Myr (right column plots in Fig. \ref{fig:3bands}), we found that the binary track nicely reproduces the $J$-band measurements; however, the tracks for this older age range seems to underestimate some of the $H$- and $K$- band measurements \citep[especially those from][]{2008A&A...482..939B}.
The most relevant result of the comparisons above is that the small disagreements in the $J$ and $H$ bands reported by \citet{2006ApJ...638..887L} and \citet{2007ApJ...665..736C} are partially alleviated (for an age of 75\,Myr), or completely removed (for an age of 120\,Myr), considering AB Dor C as a binary system. 

We also estimated the bolometric luminosity ($L_\mathrm{bol}$) using the photometric published measurements and the bolometric corrections found in \citet{2013ApJS..208....9P} and \citet{2015ApJ...810..158F}. Both corrections produce very similar results and are consistent within the errors. Conservatively we adopt the bolometric correction value of 3.10~$\pm$~0.13 from \citet{2015ApJ...810..158F} for an M7 dwarf.
Differences in bolometric luminosity or in magnitudes could highlight two scenarios. On one hand, a discrepancy in luminosity would suggest problems in the fundamental physics used to compute very low-mass stars models, in particular due to the adopted equation of state, outer boundary conditions, magnetic fields, and surface spots \citep{2001ASPC..243..581S,2007A&A...472L..17C,2010Ap&SS.328..167D,2013ApJ...779..183F,2015ApJ...807..174S,2018MNRAS.476...27T}. On the other hand, a difference visible only in magnitudes should point out for the need for more accurate synthetic spectra, which, especially in the case of low- and very low-mass stars, still represents a challenging task.
The top panels of Fig. \ref{fig:3bands} show that the binary scenario produces a slightly better agreement with the estimated bolometric luminosities. This is especially notable at the younger age of 75 Myr. Although small, this effect may be pointing to a problem in the fundamental physics used in very low-mass stars models, as previously stated.

Regarding the spectral type of AB Dor C, and according to the binary scenario, the M5.5~$\pm$~1 classification reported by \citet{2006ApJ...638..887L} should be assigned to the heavier component of the system, AB Dor Ca. To obtain an estimate of the spectral features of the weaker component Cb we used the color-magnitude calibration provided by \citet{2014A&A...568A..77Z}, in turn based on the least massive population, brown dwarfs, and giant planets, belonging to the Pleiades cluster (age 120\,Myr). Following this calibration, and for a $K_s$ magnitude of 11.7~$\pm$~0.3 (Sect. \ref{sect:results}), a spectral type of L4--6 could be expected for component AB Dor Cb.

\subsection{Orbit and stability of AB Dor Ca/Cb}

Although our interferometric measurements do not provide any information about the orbit of Cb and Ca, we can derive estimates for the semimajor axis using the conversion factors from projected separation to semimajor axis provided by \citet{2011ApJ...733..122D}. The median values given in their Table 6 for very low-mass binaries range from 0.85 to 1.16, which, considering our Ca/Cb separation (38.1 mas) and the dynamical mass of AB Dor C (0.090 M$_{\rm \odot}$), translate to a semimajor axis for the Ca/Cb orbit of 32.4--44.2 mas with a period of 418--666 days.
The Ca/Cb binary is, in turn, orbiting the 0.89 M$_\odot$ AB Dor A with a period of 
11.78 yr \citep{2017A&A...607A..10A}. It is obvious that the complete system is dynamically dominated by the presence
of AB Dor A, and accordingly, its gravitational pull exerted on the inner orbit Ca/Cb should be evaluated to ascertain if the 
latter pair is in a stable orbit.
For this purpose (and neglecting the effect of the 9" apart AB Dor Ba/Bb) we can consider AB Dor as a triple 
system (A, Ca, Cb) where AB Dor Cb is in an S-type orbit, that is, Cb is orbiting near one of the bodies (Ca) while the third body (AB Dor A) acts as a perturber. The critical semimajor axis at which the orbit of the system Ca/Cb is stable depends on the eccentricity of the binary A/C, the mass ratio A/C, and the separation between the host object (Ca) and the perturber (A).  Assuming the estimated mass of AB Dor Ca, and adopting the orbital parameters of AB Dor C around AB Dor A given in \citet{2017A&A...607A..10A}, the formulae provided by \citet{1999AJ....117..621H} yield stable orbits 
for AB Dor Ca/Cb for separations $<$50\,mas. This implies that our measured separation for the binary in AB Dor C ($\sim$38 mas) would correspond to a stable binary system. A similar conclusion can be reached following a different reasoning based on the simulations of \citet{2005A&A...434..355M}: according to their Fig. 5, stable S-type orbits 
are obtained for the estimated distance ratio ($d_{Ca-Cb}/d_{A-C}\sim$0.27) and mass ratio ($M_{C}/M_A\sim$0.10) of the triple system AB Dor A/Ca/Cb.\\

Finally, should AB Dor Cb have been detected in previous observations? 
The presence of the solar-type star AB Dor A ($K_s$ =  4.686) at only 0.2" made it extraordinarily difficult to detect and characterize AB Dor C ($K_s$ = 9.5)
\citep{2005Natur.433..286C,2007ApJ...665..736C}, given the high-contrast imaging needed in the vicinity of AB Dor A; as AB Dor Cb is about three magnitudes weaker, it is very likely that this newly discovered companion remained unnoticed in the observations reported by \citet{2007ApJ...665..736C} or \citet{2008A&A...482..939B}. 
Assuming a face-on, circular orbit and masses of 0.072~$\pm$~0.013 M$_{\rm \odot}$ and 0.013~$\pm$~0.001 M$_{\rm \odot}$ for AB Dor Ca/Cb, the radial velocity semi-amplitude produced in Ca would be $\sim$2 km/s with a period of 418--666 days. At 2.3 $\rm \mu m$ of wavelength, the expected spectral shift due to the presence of the companion Cb would be $\sim$0.02 nm while the finest spectral resolution achieved in the spectra of AB Dor C is 1.5 nm \citep{2007ApJ...665..736C} explaining why this radial velocity signal has not been discovered before.    

\section{Conclusions}

We present interferometric evidence that AB Dor C is not a single pointlike star but most likely a binary system of very low-mass objects. Our results show that both squared visibilities and closure phases are in good agreement with a binary system of $\sim$38 mas separation between the components and a K-band flux ratio of $\sim$5\%. This configuration implies masses for the tentative binary AB Dor Ca/Cb of 0.072~$\pm$~0.013 M$_{\rm \odot}$ and 0.013~$\pm$~0.001 M$_{\rm \odot}$, according to the PMS evolutionary models of \citet{2000ApJ...542..464C}, \citet{2015A&A...577A..42B} and \citet{2018MNRAS.476...27T}. It is worth noting that, with these masses, one of the objects would lie near the hydrogen-burning limit (AB Dor Ca), while AB Dor Cb would lie at the frontier between brown dwarfs and planets. 
The binarity of AB Dor Ca/Cb may have gone unnoticed in previous observations given the three-magnitude difference between Ca and Cb and the great difficulty of discerning AB Dor C itself from the nearby, five-magnitude brighter AB Dor A. However, the binary hypothesis would alleviate the disagreement between observed magnitudes and theoretical mass-luminosity relationships. We considered the two most frequently used scenarios (75$\pm$25 Myr and 120$\pm$20 Myr) and found that especially at the $J$ and $H$ bands the binary hypothesis produces a better agreement than a single 0.090 M$_{\rm \odot}$ object.

The perturbation caused by the more massive AB Dor A would destabilize a binary system of separation larger than 50 mas. With a separation of about 38 mas, the newly discovered binary AB Dor Ca/Cb appears stable under such perturbation.
Yet, although our result defines a very plausible scenario for AB Dor C, it is based on a limited number of visibilities taken near the performance limit of AMBER and, therefore, further confirmation of our findings would be convenient.
Advanced instrumentation (i.e. GRAVITY dual-field on-axis mode observations) will help to clarify the nature of this remarkable system.

\acknowledgments
Authors thank the anonymous referee for useful suggestions
that improved the manuscript. J.B.C., J.C.G., and J.M.M. were
partially supported by the Spanish MINECO projects AYA2012-
38491-C02-01, AYA2015-63939-C2-2-P, PGC2018-098915-B-
C22 and by the Generalitat Valenciana projects PROMETEO/
2009/104 and PROMETEOII/2014/057. I.M.-V. is a fellow of
the GenT program (Generalitat Valenciana) under the Project
Grant CIDEGENT 2018/021.

\bibliography{main}{}
\bibliographystyle{aasjournal}

\end{document}